**Assessment of Non-Institutional AI Tool Usage Among Clinicians**

**Sarah Pungitore and Jarrod Mosier**

**Abstract**

**Background:** Generative artificial intelligence (AI) tools are increasingly accessible and have the potential to improve efficiency across clinical workflows. However, clinicians may also use non-institutional AI tools that are not provided, managed, or governed by their healthcare institutions, creating potential concerns related to privacy, security, accuracy, and clinician-AI interaction. Little is known about how clinicians currently use these tools for work-related tasks.

**Objectives:** Characterize use of non-institutional AI tools by clinicians for work-related tasks, including the types and frequency of tools used and identify practical considerations for responsible clinician use of AI.

**Methods:** We conducted a descriptive survey of clinicians (medical students, residents, fellows, faculty, and advanced practice partitioners) recruited from the University of Arizona College of Medicine – Tucson and Banner University Medical Center – Tucson between May 20 and June 26, 2026. Participants reported their use of AI tool categories and the frequency with which they used AI for specific tasks across five workload categories: administrative work, clinical work, research, studying/continued education, and teaching.

**Results:** Forty-four respondents completed the survey. Forty-three respondents (97.7%) reported using AI for at least one work-related task during the preceding 6 months. Conversational AI and clinical decision support/diagnostic AI were the most used tool categories, each reported by 28 respondents. AI use occurred across all five workload categories and every assessed task. Administrative and clinical tasks demonstrated the most frequent use, with more than 60% of respondents reporting AI use for writing emails or messages and drafting letters, memos, or reports. AI was also used for higher-risk activities, including diagnostic assistance and clinical decision support.

**Conclusions:** Non-institutional AI use was common among surveyed clinicians and extended across a broad range of work-related activities, including tasks with potential implications for clinical reasoning and patient care. These findings highlight the need for governance frameworks that address non-institutional AI use while further research is needed to characterize how clinicians use these tools, how they evaluate AI-generated outputs, and how AI can be safely and effectively integrated into clinical workflows.

## Introduction

Generative artificial intelligence (AI) tools, such as ChatGPT,[1] have rapidly increased in popularity in recent years. Generative AI has introduced a new set of text, image, and speech-based capabilities that allow individuals to offload complex tasks to AI tools, including writing, question answering, and coding.[2] These tools are also highly accessible in a variety of formats, including web, mobile, and personal computer applications. Given these groundbreaking capabilities and widespread accessibility, AI tools have often been hailed as essential time and resource-saving solutions. In the clinical space in particular, generative AI applications to patient care, education, and research have been proposed to improve clinician efficiency, reduce healthcare costs, and improve patient outcomes.[3–9]

Despite their popularity and widely publicized benefits, recent studies and news articles have hinted at potentially serious drawbacks with the use of AI tools, such as deskilling and security risks. For example, researchers found that AI may hinder learning and skill retention, where individuals who delegated coding tasks to AI performed significantly lower on a skills assessment than those who did not use AI at all.[10] Additionally, nearly all studies assessing physician deskilling found that using AI tools reduced physician ability to make correct decisions.[11] Furthermore, there have been significant concerns related to data security and privacy. The privacy and data-use practices of widely available AI tools vary substantially by provider, product, account type, and user settings, with some consumer-facing services automatically opting user inputs into being used for model improvement.[12] Security and privacy are also frequently overlooked when using novel AI tools. For example, individuals using a popular open-source generative AI productivity tool were found to have data exposed online, including conversations and login credentials, because there was insufficient guidance on how to configure the application to prevent these breaches.[13]

Although most AI tools are not marketed specifically to clinicians, clinicians are likely using a variety of AI tools to complete work-related tasks. However, to our knowledge, few studies have been conducted to understand which tools are used, what tasks they are used for, and how frequently they are used. Additionally, there have been no studies that have assessed usage of "non-institutional" (i.e., tools that are not institutionally provided, managed, or governed) AI tools. Unlike "institutional" tools (i.e., tools that are institutionally provided, managed, and governed for specific work environments), such as medical scribes,[14] which may come with end-user license or data-use agreements, non-institutional tools like ChatGPT are typically not strictly regulated despite their widespread availability. Given the sensitive and critical nature of healthcare, understanding whether, when, and how frequently these tools are used by clinicians to complete work-related tasks is an important step towards managing AI use in clinical settings.

Additionally, as both the benefits and drawbacks of AI tools become more well-known, there is a need to better understand how clinicians are using these tools in practice and what considerations should guide their use. Recent studies aimed at clinicians have explained how machine learning and AI work,[15] described methods for selecting the best model,[16] outlined conceptual frameworks for integrating AI into clinical education and workflows,[17–21] and discussed how physician competencies may change with the introduction of AI.[22] While these studies addressed important aspects of planning for integrated healthcare AI systems, they do not address how clinicians are already using AI tools in their daily work. As a result, there remains a significant disconnect between the envisioned state of AI and the reality of how clinicians are using these tools today.

The aims of this study were thus the following:

1) Conduct a survey to characterize how clinicians use non-institutional AI tools to complete work-related tasks, including the types and frequency of AI tools used.
2) Provide practical considerations for responsible AI use by clinicians based on survey observations.
3) Highlight next steps towards understanding how clinicians and AI can optimally interact.

## Methods

### *Data Collection and Ethics*

Data collection and study methods were approved by both the University of Arizona (#STUDY00007904) and Banner Health (#UAFEAS0003280) Institutional Review Boards.

### *Study Sample*

Study participants were recruited from the University of Arizona College of Medicine – Tucson (COM-T) and Banner University Medical Center – Tucson (Banner). Individuals were included if they were medical students, residents, fellows, faculty, or advanced practice practitioners (APPs; includes nurse practitioners, physician assistants, and certified registered nurse anesthetists) employed by or enrolled at COM-T (medical students, residents, fellows, and faculty) or employed by Banner (APPs). Individuals were contacted once via a college-wide listserv at the start of the survey period (May 20, 2026 – June 26, 2026). Participants were excluded if they did not complete 100% of the survey or left all survey responses blank despite their survey being marked as fully complete.

### *Survey Details*

The full Qualtrics survey, including the informed consent statement, is included in the Supplementary Material. The survey was divided into three sections: 1) Background information; 2) AI tools used; and 3) Tasks performed using these tools. For the background information section, participants were asked about their job title (medical student, resident/fellow, APP, or faculty), primary COM-T clinical department (see Supplementary Material), and career stage (student, resident, fellow, <5 years since completing training, 5 – 10 years since completing training, and >10 years since completing training). Participants were additionally asked to indicate what percentage of their workload fell into the following categories based on the function of the work performed: Administrative Tasks, Clinical Work, Research, Studying/Continued Education, and Teaching.

For the AI tools section, we defined tool categories based on their primary function. Participants were asked to select all of the following AI tool categories they had used in the past 6 months to complete work-related tasks: 1) Conversational AI tools (Examples: ChatGPT, Claude, Google Gemini, Perplexity AI); 2) OS-Integrated AI Tools (Examples: Microsoft Copilot, Apple Intelligence); 3) Autonomous AI Agents (Examples: AutoGPT, OpenClaw, LangChain); 4) Writing and Editing Tools (Examples: Grammarly, Notion AI, Jasper AI); 5) Literature Review/Knowledge Discovery tools (Examples: Elicit, Semantic Scholar, Scite); 6) Clinical Documentation and Workflow Automation (Examples: Oracle Clinical AI Agent); 7) Clinical Decision Support/Diagnostic AI (Examples: Aidoc, OpenEvidence, Tempus AI); 8) Teaching, Tutoring, and Educational Support (Examples: Khanmigo, MagicSchool AI); 9) Media Generation (Examples: Midjourney, DALL·E, Canva, Adobe Firefly); and 10) Question Answering (Examples: Google AI Search). Participants were additionally provided a write-in option. At the time of this survey, the only institutionally available tool within COM-T and Banner was Oracle Clinical AI Agent, which was used for clinical documentation. Because participants were not specifically briefed on how to distinguish between institutional and non-institutional tools, Oracle Clinical AI Agent was not excluded from the survey to avoid confusion when categorizing AI tools. However, responses corresponding to this tool and its associated tasks were excluded from analyses of non-institutional AI tool usage when appropriate.

For the section asking participants about the tasks performed using AI tools, we generated a comprehensive list of tasks for each of the broader workload categories (Administrative Tasks, Clinical Work, Research, Studying/Continued Education, and Teaching). The list of tasks for each category are presented in Table S1; a write-in option was also provided for each category. Participants were asked to indicate how frequently they used AI tools for each task in the last six months on the following Likert scale: Never, less than once a month, once a month, 2 – 3 times a month, once a week, 2 – 3 times a week, and daily.

### *Survey Analysis*

Given the small sample size, we limited the analysis to descriptive statistics and comparisons rather than formal statistical testing. We summarized categorical variables as counts and percentages and continuous ones as medians and interquartile ranges.

## Results

### *Characteristics of Study Sample*

After applying the inclusion and exclusion criteria, there were 44 total respondents. Descriptive statistics for these individuals are presented in Table 1, stratified by AI tool usage. Most respondents were APPs or faculty who had completed training more than 10 years ago. The primary departments represented were Emergency Medicine and Medicine. Participants spent the most time on administrative and clinical work tasks. The median percentage of time spent on administrative tasks was 12.5% (Q1: 5.0%, Q3: 20.0%); clinical work was 65.0% (Q1: 23.75%, 80.0%);  research was 1.0% (Q1: 0.0%, Q3: 16.25%); studying or continued education was 1.0% (Q1: 0.0%, Q3: 6.25%); and teaching was 5.0% (Q1: 0.0%, 10.0%). We used the median percentage of time spent on each broader workload category as the cut-off to characterize differences in AI usage between respondents who spent more versus less time on each category.

### *Assessment of AI Tool Usage*

The most common tools used were Conversational AI and Clinical Decision Support/Diagnostic AI tools, each used by 28 respondents in the six months preceding the survey period, and OS-Integrated AI and Clinical Documentation and Workflow Automation tools, each used by 18 respondents in the six months preceding the survey period. The least frequently used tools were those for Teaching, Tutoring, and Educational Support, Autonomous AI Agents, Literature Review/Knowledge Discovery, and Media Generation tools, each with 3 or fewer respondents indicating usage in the six months preceding the survey period. In general, a smaller proportion of APPs used AI tools when compared to medical students, residents/fellows, and faculty across all tools surveyed, although APPs had the highest proportion of respondents using Clinical Documentation tools. We observed no clear patterns by career stage or department, with the proportion of respondents using AI varying based on the category of tools.

***Assessment of Task-Based Usage***

Of the 44 total respondents, 43 (97.7%) indicated using AI tools to complete at least one work-related task during the six months preceding the survey. Furthermore, all 43 respondents reported AI tool usage even after excluding the tool categories *Clinical Documentation* and *Workflow Automation* and the tasks *Drafting patient notes* and *Transcribing patient visits*, which were relevant to Oracle Clinical AI Agent, the only institutionally available AI tool. Figure 1 presents the number of respondents who reported using AI tools for each task, stratified by usage frequency, workload category, and time spent on each workload category, categorized as at or above the median or below the median percentage of time spent. All tasks listed in Table S1 were performed using an AI tool by at least one respondent. Administrative and clinical work tasks demonstrated the most frequent AI tool usage, with multiple participants reporting daily AI use for each task within these broader categories.

Each category also demonstrated distinct patterns of AI usage based on the amount of time respondents spent on tasks within each broader category. In general, all respondents reported using AI tools for administrative and clinical work tasks, including those who spent less than the median percentage of time performing these tasks. On the other hand, AI tools were used for research, studying or continuing education, and teaching tasks only among respondents who spent at least than the median amount of time on these activities. More specifically, for administrative tasks, over 60% of respondents reported using AI for *Writing emails or messages* and *Drafting letters, memos, or reports*. For clinical work tasks, the tasks most commonly performed with AI tools were *Drafting patient notes* and *Transcribing patient visits*. However, respondents who spent at least the median amount of time on clinical work also frequently used AI for *Diagnostic assistance* and *Clinical decision support*. For research tasks, most respondents did not use AI tools, with *Searching for or summarizing literature* being the primary exception, as 59% of respondents reported using AI for this task. For studying and continued education tasks, AI use varied by task, workload distribution, and frequency. For example, while 55% of all respondents indicated using AI tools for *Searching for educational materials*, AI tools for other tasks such as *Creating flashcards or memory aids* and *Reviewing case studies* were only used by those who spent more than the median amount of time on this category. Finally, for teaching tasks, nearly all AI use occurred among individuals who spent more than the median amount of time on teaching. The sole exception was one respondent who used AI daily for *Simplifying complex topics* and *Generating handouts* despite spending less than the median amount of time on teaching tasks.

**Discussion**

In this study, we surveyed clinicians regarding their use of non-institutional AI tools to complete work-related tasks. We characterized AI tool usage across multiple clinical departments, career stages, and workload distributions and assessed how frequently AI tools were used to complete various work-related tasks across five broader categories: Administrative, Clinical Work, Research, Studying/Continued Education, and Teaching tasks.

***Clinicians and Non-Institutional AI Use***

The primary contribution of this study was its focus on characterizing the use of non-institutional tools in clinical settings. To our knowledge, this study was among the first to systematically examine how clinicians use non-institutional tools to complete work-related tasks. From our survey results, we found AI use was widespread with 43 of 44 respondents reporting using non-institutional AI tools for at least one work-related task within the six months preceding the survey. AI use was also reported across all five broader workload categories and for every task assessed. Administrative and clinical work tasks demonstrated the most frequent AI use, with most respondents using them for writing and documentation tasks. However, AI use was also noted for potentially higher-risk clinical activities, including diagnostic assistance and clinical decision support. Thus, while AI tools were frequently used for more AI-friendly tasks like text generation and summarization, there is evidence that clinicians are also using these tools for activities that may directly influence clinical reasoning and patient care. We observed few descriptive patterns by career stage, department, or workload breakdown, suggesting that these observations were not specific to any particular subgroup we surveyed. These findings highlight the need to better understand how these tools are being used, in addition to potential benefits and drawbacks from extended usage.

***Implications for Responsible AI Use***

The observed patterns of usage in this study have several implications for the responsible use of AI in clinical workflows. Since we did not assess whether these tools are being used appropriately in this study, we developed a framework for understanding potential risks and practical considerations involved in using AI tools, both to inform clinicians today and to provide a foundation for future work in this area. We summarize these potential risks and practical considerations in Table 2. For lower risk productivity tasks, such as writing emails, clinicians can generally take a more hands-off approach of reviewing outputs for errors. However, for any task even tangentially related to patient care, clinicians must be vigilant in independently validating outputs to ensure no negative impacts on patient outcomes since AI tools can hallucinate responses. While the benefit of AI tools may be easily seen with lower risk activities, clinicians must take special care to evaluate the full range of risks and benefits for any high-risk activity and eschew AI usage if necessary. These results further underscore the need for academic and healthcare institutions to define AI policy in a way that provides sufficient governance over non-institutional AI tools. These policies should adequately address non-institutional AI tools, including expectations regarding appropriate use, privacy and security, and accountability.

***Future Directions for Clinician-AI Interaction Research***

There are several directions for future research related to the interaction of AI tools and clinicians. The first research direction is to more completely characterize how clinicians are using non-institutional tools today. Research should be conducted to understand what data is entered, how clinicians evaluate and verify AI-generated outputs, and whether there are observed short-term benefits or drawbacks from AI use, particularly related to critical thinking, judgement, and skill degradation. The second research domain is to examine the long-term impacts of AI systems, specifically answering the questions of how, why, and under what conditions AI can be safely and effectively integrated into clinician workflows. This second domain also encompasses education research, including how best to incorporate AI-specific training into medical education.

***Limitations***

There were several limitations with the methods presented in this study. First, the response rate was around 2.2%, with 44 completed responses from approximately 2,000 potential respondents across COM-T and Banner. We were unable to send follow-up requests due to college-wide email policies, which contributed to the low response rate. Additionally, individuals with greater experience with or interest in AI may have been more likely to participate, thus introducing potential self-selection bias. Furthermore, although we captured at least one response across all positions, levels of career development, and most departments, the small sample size and uneven distribution of responses prevented meaningful subgroup analyses. Second, our survey was only conducted at a single, urban medical institution and thus, our findings may not necessarily generalize to other healthcare systems or clinical settings. Third, all AI usage was self-reported and since usage was not independently verified, there may have been differences between participants in recalling AI use or in interpreting the functional categories of tools. Finally, our survey was deliberately brief to ensure as high a response rate as possible, although this design choice limited our ability to assess other valuable topics, such as the relationships between specific AI tools and the tasks for which they are used, trust in AI systems, perception of the quality of AI tools, and use of AI tools for personal tasks. These topics are all additional areas for future research.

**Conclusion**

In this study, we performed a descriptive analysis of non-institutional AI usage by surveying clinicians within a single academic medical center to determine when and how frequently these tools are used to complete work-related tasks. AI usage was common among respondents, with 44 out of 43 respondents reporting use of AI tools in the six months preceding the survey. Clinicians used AI for a variety of tasks, spanning administrative, clinical work, research, studying/continued education, and teaching activities. From these results, we identified practical considerations for responsible clinician AI use and priorities for future research to determine how clinicians and AI systems can optimally interact to improve healthcare delivery.

**References**

1. Introducing ChatGPT | OpenAI. Accessed March 3, 2026. https://openai.com/index/chatgpt/

2. Patil R, Gudivada V. A Review of Current Trends, Techniques, and Challenges in Large Language Models (LLMs). *Appl Sci*. 2024;14(5):5. doi:10.3390/app14052074

3. Cascella M, Montomoli J, Bellini V, Bignami E. Evaluating the Feasibility of ChatGPT in Healthcare: An Analysis of Multiple Clinical and Research Scenarios. *J Med Syst*. 2023;47(1):33. doi:10.1007/s10916-023-01925-4

4. Shool S, Adimi S, Saboori Amleshi R, Bitaraf E, Golpira R, Tara M. A systematic review of large language model (LLM) evaluations in clinical medicine. *BMC Med Inform Decis Mak*. 2025;25(1):117. doi:10.1186/s12911-025-02954-4

5. Lin C, Kuo CF. Roles and potential of Large language models in healthcare: A comprehensive review. *Biomed J*. 2025;48(5):100868. doi:10.1016/j.bj.2025.100868

6. Busch F, Hoffmann L, Rueger C, et al. Current applications and challenges in large language models for patient care: a systematic review. *Commun Med*. 2025;5(1):26. doi:10.1038/s43856-024-00717-2

7. Zhang K, Meng X, Yan X, et al. Revolutionizing Health Care: The Transformative Impact of Large Language Models in Medicine. *J Med Internet Res*. 2025;27(1):e59069. doi:10.2196/59069

8. Ahuja AS. The impact of artificial intelligence in medicine on the future role of the physician. *PeerJ*. 2019;7:e7702. doi:10.7717/peerj.7702

9. Matheny ME, Whicher D, Thadaney Israni S. Artificial Intelligence in Health Care: A Report From the National Academy of Medicine. *JAMA*. 2020;323(6):509-510. doi:10.1001/jama.2019.21579

10. Shen JH, Tamkin A. How AI Impacts Skill Formation. *arXiv*. Preprint posted online February 1, 2026:arXiv:2601.20245. doi:10.48550/arXiv.2601.20245

11. Heudel PE, Crochet H, Filori Q, Bachelot T, Blay JY. Artificial intelligence in medicine: a scoping review of the risk of deskilling and loss of expertise among physicians. *ESMO Real World Data Digit Oncol*. 2026;12:100693. doi:10.1016/j.esmorw.2026.100693

12. King J, Klyman K, Capstick E, Saade T, Hsieh V. User Privacy and Large Language Models: An Analysis of Frontier Developers' Privacy Policies. *Proc AAAIACM Conf AI Ethics Soc*. 2025;8(2):1465-1477. doi:10.1609/aies.v8i2.36646

13. Moltbot Molts Again And Becomes OpenClaw, Pushback And Concerns Grow. Accessed February 19, 2026. https://archive.ph/20260130152513/https://www.forbes.com/sites/ronschmelzer/2026/01/30/moltbot-molts-again-and-becomes-openclaw-pushback-and-concerns-grow/

14. Tierney AA, Gayre G, Hoberman B, et al. Ambient Artificial Intelligence Scribes: Learnings after 1 Year and over 2.5 Million Uses. *NEJM Catal*. 2025;6(5):CAT.25.0040. doi:10.1056/CAT.25.0040

15. Maiter A, Alabed S, Allen G, Alahdab F. AI in healthcare: an introduction for clinicians. *BMJ Evid-Based Med*. 2025;30(6):376-380. doi:10.1136/bmjebm-2024-112966

16. Li H, Fu JF, Python A. Implementing Large Language Models in Health Care: Clinician-Focused Review With Interactive Guideline. *J Med Internet Res*. 2025;27(1):e71916. doi:10.2196/71916

17. James CA, Wachter RM, Woolliscroft JO. Preparing Clinicians for a Clinical World Influenced by Artificial Intelligence. *J Am Med Assoc*. 2022;327(14):1333-1334. doi:10.1001/jama.2022.3580

18. Misra R, Keane PA, Hogg HDJ. How should we train clinicians for artificial intelligence in healthcare? *Future Healthc J*. 2024;11(3):100162. doi:10.1016/j.fhj.2024.100162

19. Schubert T, Oosterlinck T, Stevens RD, Maxwell PH, Schaar M van der. AI education for clinicians. *eClinicalMedicine*. 2025;79. doi:10.1016/j.eclinm.2024.102968

20. Smith H, Downer J, Ives J. Clinicians and AI use: where is the professional guidance? *J Med Ethics*. 2024;50(7):437-441. doi:10.1136/jme-2022-108831

21. Zeng F, Liang X, Chen Z. New Roles for Clinicians in the Age of Artificial Intelligence. *BIO Integr*. 2020;1:113. doi:10.15212/bioi-2020-0014

22. Schuitmaker L, Drogt J, Benders M, Jongsma K. Physicians' required competencies in AI-assisted clinical settings: a systematic review. *Br Med Bull*. 2025;153(1):ldae025. doi:10.1093/bmb/ldae025

**Table 1.** Descriptive statistics of the study sample at time of survey by the type of artificial intelligence (AI) tools used in the last six months to complete work-related tasks out of 44 completed responses. The examples provided to participants for each category were the following: Conversational AI tools (Examples: ChatGPT, Claude, Google Gemini, Perplexity AI); OS-Integrated AI Tools (Examples: Microsoft Copilot, Apple Intelligence); Autonomous AI Agents (Examples: AutoGPT, OpenClaw, LangChain); Writing and Editing Tools (Examples: Grammarly, Notion AI, Jasper AI); Literature Review/Knowledge Discovery tools (Examples: Elicit, Semantic Scholar, Scite); Clinical Documentation and Workflow Automation (Examples: Oracle Clinical AI Agent); Clinical Decision Support/Diagnostic AI (Examples: Aidoc, OpenEvidence, Tempus AI); Teaching, Tutoring, and Educational Support (Examples: Khanmigo, MagicSchool AI); Media Generation (Examples: Midjourney, DALL·E, Canva, Adobe Firefly); Question Answering (Examples: Google AI Search); and Other.

| Characteristic | Total[a] | Conversational (n=28) | OS-Integrated (n=18) | Autonomous (n=2) | Writing and Editing (n=5) | Literature Review (n=3) | Clinical Documentation (n=18) | Clinical Decision Support (n=28) | Teaching, Tutoring, and Education (n=1) | Media Generation (n=3) | Question Answering (n=20) | Other (n=1) |
|---|---|---|---|---|---|---|---|---|---|---|---|---|
| **Position n (%)** | | | | | | | | | | | | |
| APP[b] | 16 | 6 (38%) | 6 (38%) | 1 (6%) | 3 (19%) | 0 | 9 (56%) | 8 (50%) | 1 (6%) | 0 | 5 (31%) | 0 |
| Faculty | 17 | 12 (71%) | 8 (47%) | 0 | 1 (6%) | 1 (6%) | 6 (35%) | 11 (65%) | 0 | 2 (12%) | 12 (71%) | 1 (6%) |
| Medical Student | 4 | 4 (100%) | 2 (50%) | 0 | 0 | 1 (25%) | 0 | 3 (75%) | 0 | 1 (25%) | 2 (50%) | 0 |
| Resident/Fellow | 7 | 6 (86%) | 2 (29%) | 1 (14%) | 1 (14%) | 1 (14%) | 3 (43%) | 6 (86%) | 0 | 0 | 1 (14%) | 0 |
| **Career Advancement n (%)** | | | | | | | | | | | | |
| Student | 3 | 3 (100%) | 1 (33%) | 0 | 0 | 1 (33%) | 0 | 3 (100%) | 0 | 1 (33%) | 2 (67%) | 0 |
| Resident/Fellow | 7 | 6 (86%) | 2 (29%) | 1 (14%) | 1 (14%) | 1 (14%) | 3 (43%) | 6 (86%) | 0 | 0 | 1 (14%) | 0 |
| <5 years since completing training | 4 | 2 (50%) | 0 | 0 | 1 (25%) | 0 | 4 (100%) | 3 (75%) | 0 | 0 | 1 (25%) | 0 |
| 5-10 years since completing training | 7 | 3 (43%) | 3 (43%) | 0 | 2 (29%) | 0 | 3 (43%) | 4 (57%) | 1 (14%) | 1 (14%) | 3 (43%) | 0 |
| >10 years since completing training | 20 | 11 (55%) | 9 (45%) | 1 (5%) | 0 | 0 | 6 (30%) | 12 (60%) | 0 | 0 | 12 (60%) | 1 (5%) |
| **Department n (%)** | | | | | | | | | | | | |
| Emergency Medicine | 15 | 11 (73%) | 6 (40%) | 1 (7%) | 0 | 1 (7%) | 7 (47%) | 10 (67%) | 0 | 1 (7%) | 6 (40%) | 0 |
| Family and Community Medicine | 2 | 2 (100%) | 1 (50%) | 0 | 0 | 0 | 0 | 2 (100%) | 0 | 0 | 2 (100%) | 1 (50%) |
| Medicine | 11 | 7 (64%) | 5 (45%) | 0 | 2 (18%) | 1 (9%) | 3 (27%) | 9 (82%) | 1 (9%) | 1 (9%) | 4 (36%) | 0 |
| Neurology | 2 | 2 (100%) | 1 (50%) | 0 | 2 (100%) | 0 | 2 (100%) | 1 (50%) | 0 | 0 | 1 (50%) | 0 |
| Neurosurgery | 2 | 2 (100%) | 0 | 0 | 0 | 0 | 2 (100%) | 2 (100%) | 0 | 0 | 1 (100%) | 0 |
| Obstetrics and Gynecology | 1 | 1 (100%) | 1 (100%) | 0 | 0 | 0 | 1 (100%) | 1 (100%) | 0 | 0 | 0 | 0 |
| Ophthalmology and Vision Science | 1 | 0 | 0 | 0 | 0 | 0 | 0 | 0 | 0 | 0 | 1 (100%) | 0 |
| Orthopaedic Surgery | 3 | 0 | 0 | 1 (33%) | 0 | 0 | 1 (33%) | 1 (33%) | 0 | 0 | 1 (33%) | 0 |
| Pediatrics | 1 | 0 | 1 (100%) | 0 | 0 | 0 | 1 (100%) | 1 (100%) | 0 | 0 | 0 | 0 |
| Psychiatry | 1 | 1 (100%) | 1 (100%) | 0 | 0 | 0 | 0 | 0 | 0 | 0 | 1 (100%) | 0 |
| Radiology and Imaging Sciences | 1 | 1 (100%) | 1 (100%) | 0 | 1 (100%) | 1 (100%) | 1 (100%) | 0 | 0 | 1 (100%) | 1 (100%) | 0 |
| Surgery | 2 | 0 | 1 (50%) | 0 | 0 | 0 | 0 | 1 (50%) | 0 | 0 | 1 (50%) | 0 |
| Urology | 2 | 1 (50%) | 0 | 0 | 0 | 0 | 0 | 0 | 0 | 0 | 1 (50%) | 0 |

[a]Individuals may not sum to total due to missing responses.
[b]APP: Advanced Practice Provider. Includes nurse practitioners, physician assistants, and certified registered nurse anesthetists.

**Table 2.** Potential risk and practical considerations for AI use for all tasks assessed.

| Category | Task | Potential Risk | Practical Considerations |
|---|---|---|---|
| **Administrative** | Writing emails or messages | Low | Consider whether AI provides meaningful benefit and review responses for errors. |
| | Drafting letters, memos, or reports | Low | Review content for factual or contextual errors and ensure sensitive information is not unnecessarily entered. |
| | Automating repetitive tasks | Low | Confirm automation performs the intended task correctly and does not expose sensitive information or introduce unintended changes. |
| | Scheduling or calendar assistance | Low | Verify scheduling details before acting on AI-generated outputs. |
| | Documentation cleanup or formatting | Low | Review final document for accuracy. |
| | Creating templates or standardized forms | Low | Review templates for accuracy, appropriateness, and compliance with organizational requirements before use. |
| | Data entry or spreadsheet assistance | Medium | Verify data accuracy, particularly for data used for downstream decisions. |
| | Policy or procedure drafting | Medium | Review outputs for accuracy and compliance with institutional policies and procedures. |
| | Meeting agenda or minutes preparation | Low | Review outputs for accuracy. |
| **Clinical Work** | Drafting patient notes | High | Ensure patient information is authorized for use with the tool and review for inaccuracies before incorporation into the medical record. |
| | Explaining diagnoses, procedures, or treatment plans | High | Independently verify clinical information and ensure explanations are appropriate for the specific patient and clinical context. |
| | Transcribing patient visits | High | Ensure patient information is appropriately protected and the tool is authorized for clinical documentation. Review transcripts for accuracy. |
| | Answering patient questions | High | Review responses for accuracy and appropriateness before communicating them to patients. |
| | Diagnostic assistance | High | Treat AI output as decision support rather than a definitive diagnosis and independently verify recommendations. |
| | Evidence review for clinical questions | High | Verify cited evidence and ensure summaries accurately reflect the underlying literature. |
| | Clinical decision support | High | Independently verify AI recommendations and maintain clinician responsibility for the final decision. |
| | Medication-related support | High | Independently verify all AI-provided information. |
| | Patient education material creation | High | Review content for accuracy and appropriateness before distribution. |
| **Research** | Brainstorming research questions or hypotheses | Low | Treat AI-generated ideas as suggestions and independently assess their scientific merit. |
| | Searching for or summarizing literature | Medium | Verify content from all sources while being mindful of fabricated or misrepresented references. |
| | Assisting with protocol or IRB draft text | Medium | Review content for accuracy and ensure that sensitive study information is handled appropriately. |
| | Drafting or editing manuscripts | Medium | Independently verify factual claims and citations and ensure appropriate interpretation and wording. |
| | Writing or editing grant applications | Medium | Review AI-generated content for accuracy and ensure confidential proposal information is appropriately protected. |
| | Reviewing statistical or methodological approaches | High | Independently evaluate AI recommendations with experts before incorporating them into research. |
| | Generating code or analysis outlines | Medium | Review and test generated code for appropriate implementation of the intended analysis. |
| **Studying or Continued Education** | Searching for educational materials | Low | Verify the credibility and currency of recommended resources. |
| | Summarizing educational materials | Low | Compare summaries with the original material when accuracy is important and add missing context as appropriate. |
| | Creating study guides or outlines | Low | Review content for accuracy and use authoritative educational resources as appropriate. |
| | Generating practice questions or quizzes | Low | Verify questions and answers for factual accuracy and alignment with the intended learning objectives. |
| | Explaining difficult concepts | Medium | Independently verify explanations, particularly for clinically relevant or technically complex concepts. |
| | Comparing clinical guidelines or recommendations | Medium | Verify comparisons against current guidelines and ensure differences are accurately characterized. |
| | Preparing for exams or certifications | Low | Use AI as a supplemental learning resource and verify information. |
| | Creating flashcards or memory aids | Low | Review generated content for accuracy before incorporating into study materials. |
| | Reviewing case studies | Medium | Verify clinical reasoning and conclusions rather than relying solely on AI-generated interpretations. |
| **Teaching** | Generating teaching materials | Medium | Review materials for accuracy and alignment with learning objectives. |
| | Generating images | Low | Verify accuracy and appropriateness of generated images. |
| | Suggesting outlines | Low | Review the proposed structure for completeness and alignment with teaching objectives. |
| | Simplifying complex topics | Medium | Ensure that simplification does not introduce inaccuracies or omit nuance. |
| | Generating handouts | Medium | Review content for accuracy. |
| | Summarizing lecture content | Low | Verify accuracy of summaries against the original lecture. |

**Figure 1.** Count of respondents who indicated using generative artificial intelligence (AI) tools in the past six months for each task. The number associated with each category refers to the number of individuals who indicated using an AI tool for at least one of the tasks within the broader category. The frequency tags are the following: A (Never); B (Less than once a month); C (Once a month); D (2 – 3 times a month); E (Once a week); F (2 – 3 times a week); and G (Daily). The first bar for each task is more transparent to emphasize that it is the *Never* category. Each bar is divided into respondents whose workload allocation for that broader category was at or above the median (bottom section) versus below the median (top section).

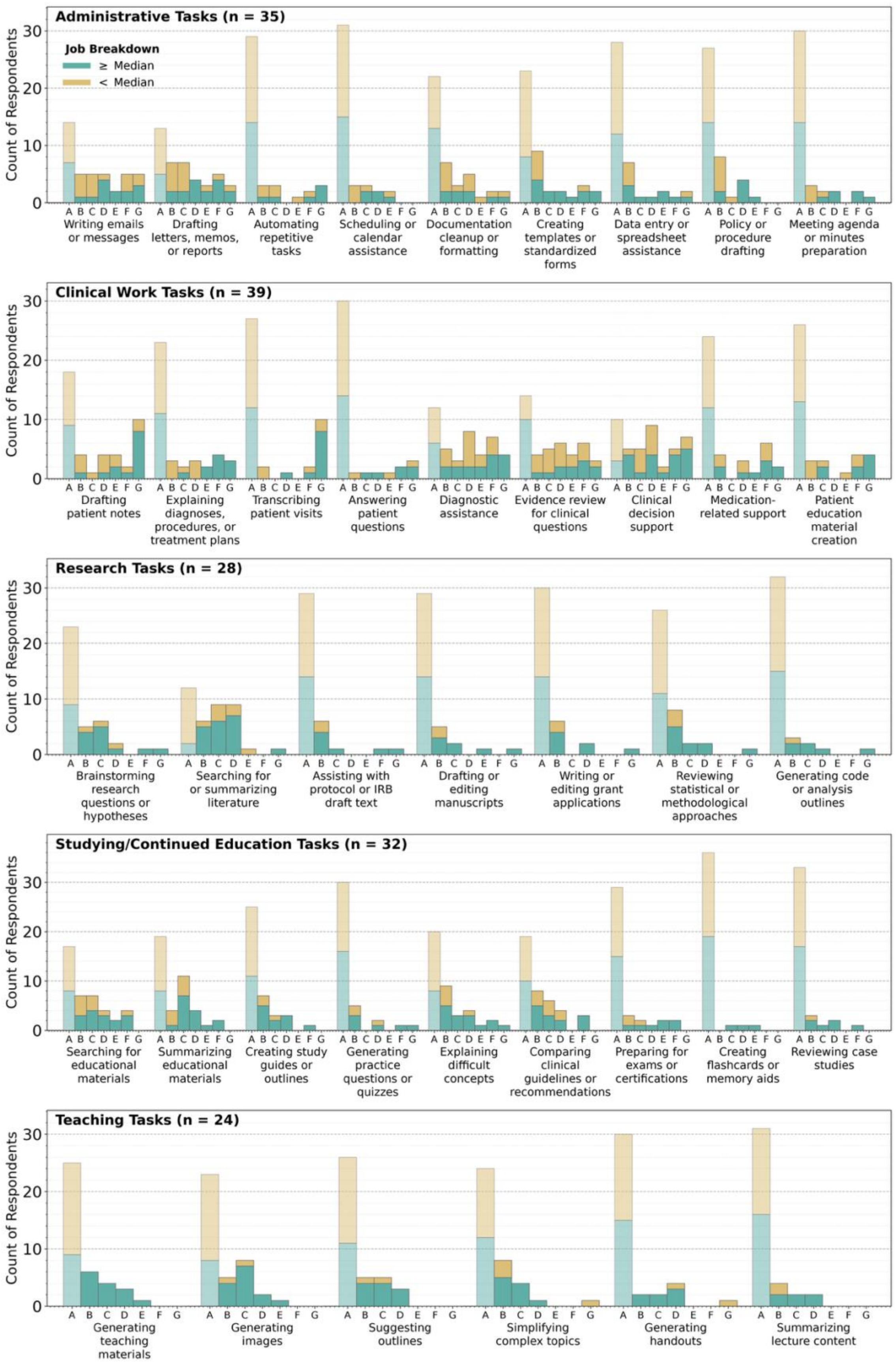

Administrative Tasks (n = 35)
Job Breakdown
≥ Median
< Median
Count of Respondents
Writing emails or messages
Drafting letters, memos, or reports
Automating repetitive tasks
Scheduling or calendar assistance
Documentation cleanup or formatting
Creating templates or standardized forms
Data entry or spreadsheet assistance
Policy or procedure drafting
Meeting agenda or minutes preparation
Clinical Work Tasks (n = 39)
Drafting patient notes
Explaining diagnoses, procedures, or treatment plans
Transcribing patient visits
Answering patient questions
Diagnostic assistance
Evidence review for clinical questions
Clinical decision support
Medication-related support
Patient education material creation
Research Tasks (n = 28)
Brainstorming research questions or hypotheses
Searching for or summarizing literature
Assisting with protocol or IRB draft text
Drafting or editing manuscripts
Writing or editing grant applications
Reviewing statistical or methodological approaches
Generating code or analysis outlines
Studying/Continued Education Tasks (n = 32)
Searching for educational materials
Summarizing educational materials
Creating study guides or outlines
Generating practice questions or quizzes
Explaining difficult concepts
Comparing clinical guidelines or recommendations
Preparing for exams or certifications
Creating flashcards or memory aids
Reviewing case studies
Teaching Tasks (n = 24)
Generating teaching materials
Generating images
Suggesting outlines
Simplifying complex topics
Generating handouts
Summarizing lecture content